%% file: lolitha.tex
\documentclass[12pt]{article}

\input preamble.tex
\title{Machine Learning for Malware Evolution Detection}

\author{Lolitha Sresta Tupadha\footnotemark[1]\ \ \ 
Mark Stamp\footnotemark[1]\,\,\footnotemark[2]}

\begin{document}

\symbolfootnotetext[1]{Department of Computer Science, San Jose State University}
\symbolfootnotetext[2]{mark.stamp$@$sjsu.edu}

\maketitle

\abstract
Malware evolves over time and antivirus must adapt to such evolution. 
Hence, it is critical to detect those points in time where malware has evolved 
so that appropriate countermeasures can be undertaken. In this research, 
we perform a variety of experiments on a significant number of malware families
to determine when malware evolution 
is likely to have occurred. All of the evolution detection techniques that 
we consider are based on machine learning and can be fully automated---in 
particular, no reverse engineering or other labor-intensive manual analysis 
is required. Specifically, we consider analysis based on hidden Markov models (HMM)
and the word embedding techniques HMM2Vec and Word2Vec.

\section{Introduction}
\label{chap:introduction}

Malware is software that is intended to be malicious
in its effect~\cite{aycock2006computer}.
By one recent estimate, there are more than one billion malware programs
in existence, with~560,000 new malware samples discovered 
every day~\cite{MercaldoFrancesco2018Aeso}. Clearly, malware is 
a major cybersecurity threat, if not the most serious security threat today.

Since the creation of the ARPANET in~1969, 
there has been an exponential growth in the number of users of the Internet. 
The widespread use of computer systems along with continuous Internet connectivity 
of the ``always on'' paradigm makes modern computer systems prime 
targets for malware attacks. 
Malware comes in many forms, including viruses, worms, backdoors, 
trojans, adware, ransomware, and so on. Malware is a continuously evolving 
threat to information security. 

In the field of malware detection, a signature typically consists 
of a string of bits that is present in a malware executable. 
Signature-based detection is the most popular method of malware detection used by 
anti-virus (AV) software~\cite{aycock2006computer}. But malware has become increasingly 
difficult to detect with standard signature-based approaches~\cite{StampMark2018Itml}. 
Virus writers have developed advanced metamorphic generators and obfuscation techniques 
that enable their malware to easily evade signature detection. For example, in~\cite{borello2008code}, 
the authors prove that carefully constructed metamorphic malware can successfully 
evade signature detection. 

Koobface is a recent example of an advanced form of malware. 
This malware was designed to target the users of social media, and its infection 
is spread via spam that is sent through social networking 
websites. Once a system is infected, Koobface gathers a user's sensitive information 
such as banking credentials, and it blocks the user from accessing 
anti-virus or other security-related websites~\cite{koob}.

Malware writers modify their code to deal with advances in detection, 
as well as to add new features to existing malware~\cite{BaratMarius2013Asoc}. 
Hence, malware can be perceived as evolving over time. 
To date, most research into malware evolution has relied on software
reverse engineering~\cite{GuptaA2009Aeso}, which is labor intensive. 
Our goal is to detect malware evolution automatically, 
using machine learning techniques. We want to find points in time where it is 
likely that significant evolution 
has occurred within a given malware family. It is important to detect such 
evolution, as these points are precisely where
modifications to existing detection strategies are urgently needed.

We note in passing that malware evolution detection can play an additional
crucial role in malware research, beyond updating existing detection strategies
to deal with new variants. Generally, in malware research, we consider
samples from a specific family, without regard to any evolutionary changes
that may have occurred over time. An adverse side effect of such an 
approach is that---with respect
to any specific point in time---we are mixing together past, present, and
future samples. Relying on training based on future samples to detect past (or present) 
samples is an impossibility in any real-world setting, yet it is seldom accounted 
for in research. By including an accurate evolutionary timeline, we can
conduct far more realistic research. Thus, accurate information regarding
malware evolution will also serve to make
research results more realistic and trustworthy.

We consider several machine learning techniques to identify potential malware evolution,
and our experiments are conducted using a significant number of malware families
containing a large numbers of samples collected over an extended period of time. 
We extract the opcode sequence from each malware sample, and these sequences are
used as features in our experiments. We group the available samples based on time 
periods and we train machine learning models on time windows.
We compare the models to determine likely evolutionary points---substantial differences 
in models across a time boundary indicate significant change in the 
code base of the malware family under consideration. 
Specifically, we experiment with
hidden Markov models (HMM) and word embedding techniques
(Word2Vec and HMM2Vec). For comparison, we also 
consider logistic regression.

The remainder of this paper is organized as follows. 
In Section~\ref{chap:back}, we discuss a range of relevant background
topics, including malware, related work, our dataset, and we introduce
the learning techniques that we employ in our experiments. 
Section~\ref{chap:results} 
contains our the experimental results, while Section~\ref{chap:conclusion} 
gives our conclusions along with a discussion of a few potential avenues for future work.

\section{Background}\label{chap:back}

In this section, we first give a brief introduction to malware. 
Then we consider related work in the area of malware
evolution detection. 

\subsection{Malware}

A computer worm is a kind of malware that spreads by itself over 
a network~\cite{aycock2006computer}. 
Examples of famous worms include Code Red, Blaster, Stuxnet, Santy,
and, of course, the Morris Worm~\cite{stamp2011information}.

Viruses are the most common form of malware, and the word ``virus'' is often used
interchangeably with ``malware.'' A computer virus is similar to a worm but it 
requires outside assistance to transmit its infection from one system to another. 
Viruses are often considered to be parasitic, in the sense that they embed themselves
in benign code. 
More advanced forms of viruses (and malware, in general) often use encryption,
polymorphism, or metamorphism as means to evade detection~\cite{aycock2006computer}.
These techniques are primarily aimed at defeating signature-based detection,
although they can also be effective against more advanced detection strategies.

A trojan horse, or simple a trojan, is malicious software that appears to be innocent 
but carries a malicious payload. Trojans are particularly popular today, with the the vast majority
of Android malware, for example, being trojans.

A trapdoor or backdoor is malware that allows unauthorized access to an infected 
system~\cite{stamp2011information}. Such access allows an attacker to
use the system in a denial of service (DoS) attack, for example.


Traditionally, malware detection has relied on static signatures, 
which typically consist of strings of bits found in specific malware samples.
While effective, signatures can be defeated by a wide variety of 
obfuscation and morphing techniques, and the sheer number
of malware samples today can make signature scanning infeasible.

Recently, machine learning and deep learning techniques have become the tools of
choice for malware detection, classification, and analysis. 
We would argue that it is also critical to detect malware evolution, since
we need to know when a malware family has evolved in a significant way
so that we can update our detection techniques to account for such changes.
As we see in the next section, this aspect of malware analysis has, 
thus far, received only limited attention
from the research community.

\subsection{Related Work}\label{chap:background}

While there is a great deal of research involving applications of machine learning 
to malware detection, classification, and analysis, there are very few
articles that consider malware evolution. In~\cite{ma2006finding}, 
analysis of malware based on code injection is considered. 
This works deals with shell code extracted from malware samples. 
The researchers used clustering techniques to analyze shell code 
to determine relationships between various samples. This work was 
successful in determining the similarities between samples, showing 
that a significant amount of code sharing had occurred. A drawback 
to the approach in this paper is that the authors only considered analysis 
of shell code. While shell code often serves as the attack vector for malware,
other attack vectors are possible, and malware evolution is not restricted
to the attack portion of the code. For example, a malware family might evolve
to be more stealthy or obfuscated, without affecting the attack payload.
Another limitation of this research is that it only considers software
similarity, and not malware evolution, per se.

Malware evolution research is considered in~\cite{gupta2009empirical}. 
One positive aspect of this research is that it considers a large dataset 
that spans two decades. The authors use techniques based on graph pruning
and they claim to show specific properties of various families are inherited from 
other families. However, it is not clear whether these properties are inherited 
from other families, or were developed independently. In addition, this work 
relies on manual investigation. A primary goal of our research is to 
eliminate the need for such manual intervention.
			
The research presented in~\cite{ouellette2013countering} is focused on detecting malware 
variants, which can be considered as a form of evolution detection. The authors apply
semi-supervised learning techniques to malware samples that have been shown to evade 
machine learning based detection. In contrast, in our research, we use unsupervised learning 
techniques to detect significant evolutionary points in time which, again, serves
to minimize the need for manual intervention.

The authors of~\cite{MercaldoFrancesco2018Aeso} extract variety of features from 
Android malware samples, and then determine various trends based standard 
software quality metrics. These results are then compared to trends present 
in Android goodware. This work shows that the trends in the Android malware 
and goodware are similar, with changes in malware following a similar path 
as goodware. These results are not surprising, given that Android malware 
largely consists of trojans that, by necessity, would tend to have a great deal
of overlap with goodware.
										
The work presented in~\cite{ChenZhongqiang2012Mcat} is focussed on malware taxonomy,
which provides some insights into malware evolution, in the form of genealogical trajectories. 
This research is based on features extracted from malware encyclopedia entries,
which have been developed by antivirus software vendors, such as TrendMicro. 
The authors use SVMs and language processing techniques to extract 
features on which their results are based.
					
In general, the features used in malware analysis can be considered to be either
static or dynamic. Static features are those that can be collected without executing 
the code, whereas dynamic features require code execution or emulation. 
In general, static features are easier to collect, while dynamic features are more 
robust with respect to common obfuscation techniques~\cite{DamodaranAnusha2017Acos}.

The authors of~\cite{rezaei2016malware} use multiple static features to perform malware 
classification among various families. The static features that are considered 
are byte $n$-grams, entropy, and image representations. In addition, hex-dump 
based features are also used, along with features extracted from disassembled files, including
opcodes, API calls, and sectional information from portable executable (PE) files. 
This works provides interesting insights on a wide variety of static features.

The research that we present in this paper can be viewed as a continuation of work 
that originated in~\cite{wadkar2020detecting}, where static PE file features of malware 
samples are used as the basis for malware evolution detection. This previous research 
employed linear support vector machine (SVM) techniques to train on samples from a
specific family over sliding windows of time. The resulting SVM weights 
are compared based on a~$\chi^{2}$ measure, and observed differences in model 
weights are used to indicate potential evolutionary points in time.

The work in~\cite{paul2021word}, which employs opcode sequences from malware 
samples to analyze malware evolution, is related to the research presented 
in~\cite{wadkar2020detecting}. In~\cite{paul2021word}, the  
data is again divided into time windows, and support vector machine (SVM) techniques 
are used to observe evolutionary points in the malware samples. In addition, 
hidden Markov model (HMM) techniques are used as a secondary test 
to confirm suspected evolutionary points in time. Our research in this paper 
is a further extension to this previous work. 
We perform extensive experiments with HMMs and
the word embedding techniques of Word2Vec and HMM2Vec 
to analyze malware evolution. We find that we can automatically detect
significant evolution in malware families using these techniques.

\subsection{Dataset}

The dataset we use in this research consists of Windows portable executable files 
belonging to~15 malware families. Two families (Winwebsec and Zbot) 
are from the Malicia dataset~\cite{nappa2015malicia}, while the remaining
families are from a larger dataset that was constructed using 
VirusShare~\cite{KimSamuel2018PHAf}. Each malware family contains a
a number of samples from an extended period of time. 
Samples belonging to a malware family are assumed to have similar characteristics 
and to share a code base. However, samples within the same family 
differ, as malware writers regularly modify successful malware to
perform slightly different functions, to make it harder to detect, or for
other purposes. The number of samples in each family in our dataset
is given in Table~\ref{tab:1}. 
The table also includes the time range over
which the samples were produced.

\begin{table}[!htb]
    \caption{Number of samples used in experiments}\label{tab:1}
    \centering
    \adjustbox{scale=0.85}{
    \begin{tabular}{c|cc}
    \midrule\midrule
         \textbf {Family} &  \textbf {Samples} &  \textbf {Years} \\
         \midrule
        Adload & \z791 & 2009\textendash2011\\
        Bho & 1116  & 2007\textendash2011\\
        Bifrose & \z577 & 2009\textendash2011\\
        CeeInject & \z742 & 2009\textendash2012\\
        DelfInject & \z401 & 2009\textendash2012\\
        Dorkbot & \z222 & 2005\textendash2012\\
        Hupigon & \z449 & 2009\textendash2011\\
        Ircbot & \z\z59 & 2009\textendash2012\\
        Obfuscator & \z670 & 2004\textendash2017\\
        Rbot & \z127 & 2001\textendash2012\\
        Vbinject & 2331 & 2009\textendash2018\\
        Vobfus & \z700 & 2009\textendash2011\\
        Winwebsec & 1511 & 2008\textendash2012\\
        Zbot & \z835 & 2009\textendash2012\\
        Zegost & \z506 & 2008\textendash2011\\
        \midrule
        Total &  11,037\z & 2001\textendash2018 \\
        \midrule\midrule
    \end{tabular}
    }
\end{table}


The malware families in our dataset encompass a wide variety of types, 
including virus, trojan, backdoor, worms, and so on. Some of the families 
uses encryption and other obfuscation techniques in an effort to evade detection.
Next, we briefly discuss each of the malware families listed in Table~\ref{tab:1}.

\begin{description}

\item\textbf{Bifrose} is a backdoor trojan~\cite{Bifrose}. 
As mentioned above,
a trojan poses as innocent software to trick the user into installing it,
while a backdoor serves to give an attacker unauthorized access
to an infected system.

\item\textbf{CeeInject} performs various malicious operations. 
CeeInject uses obfuscation techniques to evade signature detection~\cite{CeeInject}.

\item\textbf{DelfInject} is a worm that resides on websites 
and is downloaded to a user’s machine when visiting an infected site. 
This malware is executed whenever the system is restarted~\cite{DelfInject}. 

\item\textbf{Dorkbot} is a worm that is used to steal credentials of users on an infected system. It 
performs a denial of service (DoS) attack, and it is spread via messaging applications~\cite{Dorkbot}.

\item\textbf{Hotbar} is an adware virus that resides on websites and is downloaded 
onto a user's system when visiting a site that hosts the malware. 
Hotbar is more annoying than harmful, as it displays advertisements 
when the user browses the Internet~\cite{Hotbar}.

\item\textbf{Hupigon} is also a backdoor trojan, similar to Bifrose~\cite{Hupigon}.

\item\textbf{Obfuscator} evades signature detection using sophisticated obfuscation techniques. 
It can perform a variety of malicious activities~\cite{Obfuscator}.

\item\textbf{Rbot} is a backdoor trojan that allows attackers into the system through 
an IRC channel. This is a relatively advanced malware that is typically used to launch 
denial of service (DoS) attacks~\cite{Rbot}.

\item\textbf{VbInject} uses encryption techniques to evade signature detection. 
Its primary purpose is to disguise other malware that can be hidden inside 
of it. Its payload can vary from harmless to severe~\cite{VBinject}.

\item\textbf{Vobfus} is a trapdoor that lets other malware into the system. 
It exploits the vulnerabilities of the Windows operating system
\texttt{autorun} feature to spread on a network. This malware makes
changes to the system configuration that cannot be easily undone~\cite{Vobfus}.

\item\textbf{Winwebsec} is a trojan that attempts to trick a user into paying money 
by portraying itself as anti-virus software. It gives deceptive messages 
claiming that the system has been infected~\cite{Winwebsec}.

\item\textbf{Zbot} is a trojan that steals private
user information from an infected system. It can target information such as
system data and banking details, and it can  be easily modified to acquire 
other types of data. This trojan is generally spread via spam~\cite{Zbot}.

\item\textbf{Zegost} is another backdoor trojan that gives an attacker access 
to a compromised system~\cite{Zegost}.

\end{description}


We obtain Windows PE files for each sample in the families discussed above. 
All of our analysis is based on opcodes, so we first disassemble the files and
extract the mnemonic opcode sequence from each, discarding labels, directives, and so on. 
Since opcodes encapsulate the function of the program
we can expect opcode sequences to be useful in detecting code evolution. 
The resulting opcode sequence will serve as input to our machine learning techniques.
In addition, we segregate the samples from each family according to their creation date. 
Next, we briefly describe each of the learning techniques considered in this paper.

\subsection{Learning Techniques}

In this section, we discuss the learning techniques that are used in our
experiments. Specifically, we introduce hidden Markov models, HMM2Vec,
Word2Vec, and logistic regression.

\subsubsection{Hidden Markov Models}

\begin{figure}[!htb]
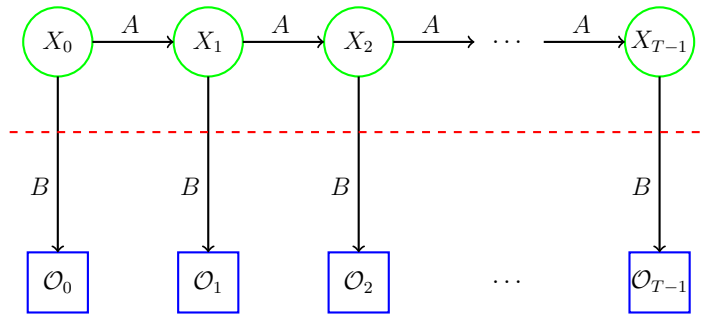

    \centering
    \input figures/hmm.tex
    \caption{Hidden Markov model~\cite{MarkStamp}}\label{fig:2}
\end{figure}

As the name suggests, a hidden Markov model (HMM) includes a Markov process that
is ``hidden'' in the sense that it cannot be directly observed. We do have access to 
a series of observations that are probabilistically related to the underlying (hidden)
Markov process. We can train a model
to fit a given observation sequence and, given a model, we can score
an observation sequence to determine how closely it fits the model.
A generic HMM is illustrated in Figure~\ref{fig:2}.	

The number of hidden states in an HMM is denoted as~$N$, and hence~$A$
in Figure~\ref{fig:2} is an~$N\times N$ row stochastic matrix that drives
the hidden Markov process. The number of distinct observation symbols
is denoted as~$M$. The~$B$ matrix in Figure~\ref{fig:2} is~$N\times M$,
with each row representing a discrete probability distribution on the 
symbols, relative to a given (hidden) state. The~$B$ matrix serves to (probabilistically)
relate the hidden states to the observations. Note that the~$B$ matrix is also
row stochastic. An HMM is specified as~$\lambda=(A,B,\pi)$, where~$\pi$
is a~$1\times N$ initial state distribution matrix.

\subsubsection{Word2Vec}\label{sect:w2v}

Word2Vec is a technique for embedding terms in a
high-dimensional space, where the term embeddings
are obtained by training a shallow neural network.
After the training process, 
words that are more similar in context will tend to be 
closer together in the Word2Vec space.

Perhaps surprisingly, meaningful algebraic properties hold for Word2Vec
embeddings. For example, according to~\cite{w2v}, if we let
$$
  w_0=\mbox{``king''}, w_1=\mbox{``man''}, w_2=\mbox{``woman''}, w_3=\mbox{``queen''}
$$
and~$V(w_i)$ is the Word2Vec embedding of word~$w_i$, then~$V(w_3)$
is the vector that is closest---in terms of cosine similarity---to
$$
  V(w_0) - V(w_1) + V(w_2)
$$

Suppose that we have a vocabulary of size~$M$.
We can encode each word as a ``one-hot'' vector
of length~$M$. For example, suppose that our vocabulary consists of
the set of~$M=8$ words
\begin{align*}
   W &= (w_0,w_1,w_2,w_3,w_4,w_5,w_6,w_7) \\
        &= (\mbox{``for''}, \mbox{``giant''}, \mbox{``leap''}, \mbox{``man''}, \mbox{``mankind''},
   		\mbox{``one''}, \mbox{``small''}, \mbox{``step''})   	   
\end{align*}
Then we encode ``for'' and ``man'' as
$$
  E(w_0)=E(\mbox{``for''}) = 10000000 
  	\mbox{\ \ and\ \ } E(w_3)=E(\mbox{``man''}) = 00010000
$$
respectively.

Now, suppose that our training data consists of the phrase
\begin{equation}\label{eq:oneSmall}
  \mbox{``one small step for man one giant leap for mankind''}
\end{equation}
To obtain training samples, we specify a window size, and for each offset 
we use all pairs of words within the specified window. 
For example, if we select a window
size of two, then from~\eref{eq:oneSmall},
we obtain the training pairs in Table~\ref{tab:train_Word2Vec}.

\begin{table}[!htb]
  \caption{Training data}\label{tab:train_Word2Vec}
  \centering\def\vp{\vphantom{step}}
  \adjustbox{scale=0.85}{
  \begin{tabular}{l|l} \midrule\midrule
    \multicolumn{1}{c|}{\textbf{Offset}\hspace*{0.5in}} 
    	& \multicolumn{1}{c}{\textbf{Training pairs}\hspace*{0.8in}} \\ \midrule
``\fbox{one\vp} small step $\ldots$'' &   (one,small), (one,step) \\
``one \fbox{small\vp} step for $\ldots$'' &  (small,one), (small,step), (small,for) \\
``one small \fbox{step\vp} for man $\ldots$'' &   (step,one), (step,small), (step,for), (step,man) \\
``$\ldots$ small step \fbox{for\vp} man one $\ldots$'' & (for,small), (for,step), (for,man), (for,one) \\
``$\ldots$ step for \fbox{man\vp} one giant $\ldots$'' & (man,step), (man,for), (man,one), (man,giant) \\
``$\ldots$ for man \fbox{one\vp} giant leap $\ldots$'' &  (one,for), (one,man), (one,giant), (one,leap) \\
``$\ldots$ man one \fbox{giant\vp} leap for $\ldots$'' & (giant,man), (giant,one), (giant,leap), (giant,for) \\
``$\ldots$ one giant \fbox{leap\vp} for mankind'' & (leap,one), (leap,giant), (leap,for), (leap,mankind) \\
``$\ldots$ giant leap \fbox{for\vp} mankind'' & (for,giant), (for,leap), (for,mankind) \\
``$\ldots$ leap for \fbox{mankind\vp}'' & (mankind,leap), (mankind,for) \\
  \midrule\midrule
  \end{tabular}
  }
\end{table}

Consider the pair ``(for,man)'' from the fourth row in 
Table~\ref{tab:train_Word2Vec}. As one-hot vectors, 
this training pair corresponds to input~10000000 and output~00010000.

\begin{figure}[!htb]
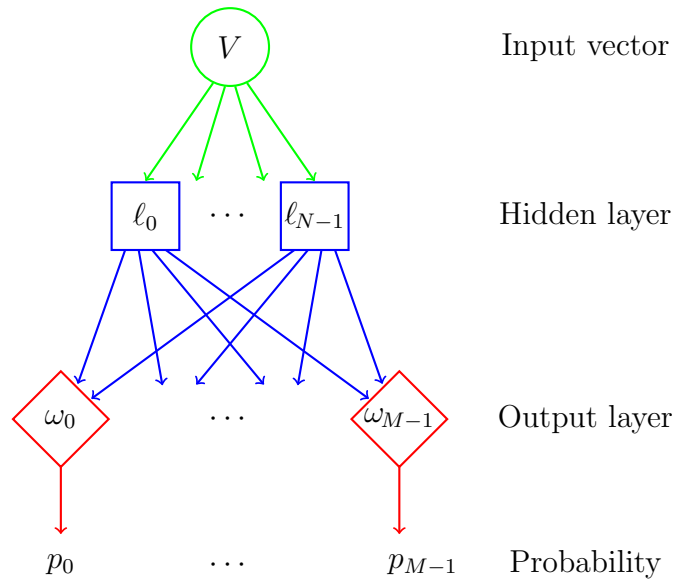

  \centering
    \input figures/w2v.tex
  \caption{Neural network for generating Word2Vec embeddings}\label{fig:w2v}
\end{figure}

A neural network similar to that in Figure~\ref{fig:w2v} is used to generate
Word2Vec embeddings. The input is a one-hot vector of length~$M$
representing the first element of a training pair, such as those in 
Table~\ref{tab:train_Word2Vec}, and the network is trained to output
the second element of the ordered pair. The hidden layer consists of~$N$
linear neurons and the output layer uses a softmax function to generate~$M$
probabilities, where~$p_i$ is the probability of the output vector corresponding to~$w_i$
for the given input. 

Observe that the Word2Vec network in Figure~\ref{fig:w2v}
has~$N\!M$ weights that are to be determined,
as represented by the blue lines from the hidden layer to the 
output layer. For each output
node~$\omega_i$, there are~$N$ edges (i.e., weights) from the hidden layer.
The~$N$ weights that connect to output node~$\omega_i$ form 
the Word2Vec embedding~$V(w_i)$ of the word~$w_i$. 

A Word2Vec model can be trained using either a
continuous bag-of-words (CBOW) or a skip-gram model. 
The model discussed in this section uses the CBOW approach, and that
is what we employ in our experiments in this paper. Note that
in our implementation, we use opcodes as the ``words.''

Several tricks are used to speed up the training of Word2Vec models.
Such details are beyond the scope of this paper; see~\cite{w2v2} for more information.

\subsubsection{HMM2Vec}\label{sect:HMM2Vec}

Analogous to Word2Vec, we can use the~$B$ matrix of a trained HMM to specify vector
embeddings corresponding to the observations. More precisely, each column of the~$B$
matrix is associated with a specific observation, and hence we obtain vector embeddings
of length~$N$ directly from the~$B$ matrix---we refer to the resulting embedding as HMM2Vec.
Since HMM2Vec is not a standard vector embedding technique,
in this section, we illustrate the process using a simple English text example.

Recall that an HMM is defined by the three matrices~$A$, $B$, and~$\pi$, 
and is denoted as~$\lambda=(A,B,\pi)$. The~$\pi$ matrix contains
the initial state probabilities, $A$
contains the hidden state transition probabilities, 
and~$B$ consists of the observation 
probability distributions corresponding to the hidden states.
Each of these matrices is row stochastic, that is, each row satisfies the
requirements of a discrete probability distribution.
Notation-wise, we let~$N$ be the number of hidden states,
$M$ is the number of distinct observation symbols, and~$T$ is the length
of the observation (i.e., training) sequence. 
Note that~$M$ and~$T$ are determined by the training data,
while~$N$ is a user-defined parameter. For more details in HMMs,
see~\cite{MarkStamp} or Rabiner's fine tutorial~\cite{LRR}.

Suppose that we train an HMM on a sequence of letters extracted from English text,
where we convert all upper-case letters to lower-case, and we discard any character that is
not an alphabetic letter or word-space. Then~$M=27$, and we select~$N=2$ hidden states,
and suppose we use~$T=50{,}000$ observations for training. 
Note that each observation is one of the~$M=27$
symbols (letters, together with word-space). For the example discussed below,
the sequence of~$T=50{,}000$ observations was obtained from 
the Brown corpus of English~\cite{BrownCorpus}, but
any source of English text could be used.

For one specific case, an HMM trained with the parameters listed in the previous
paragraph yields the~$B$ matrix in Table~\ref{tab:initFinal_B}.
Observe that this~$B$ matrix gives us two probability distributions over the observation
symbols---one for each of the hidden states. We observe that one hidden state
essentially corresponds to vowels, while the other corresponds to consonants.
This simple example nicely illustrates the machine learning aspect of HMMs,
as no a priori assumption was made concerning consonants and vowels,
and the only parameter we selected was the number of hidden states~$N$.
The training process enabled the model to learn a crucial aspect of English 
directly from the data.

\begin{table}[!htb]
  \caption{Final $B^{\transpose}$ for HMM}\label{tab:initFinal_B}
  \centering
  \adjustbox{scale=0.85}{
  \begin{tabular}{c|cc||c|cc} \midrule\midrule
\multirow{2}{*}{\textbf{Letter}} & \multicolumn{2}{c||}{\textbf{State}} 
	& \multirow{2}{*}{\textbf{Letter}} & \multicolumn{2}{c}{\textbf{State}} \\
 & 0 & 1 & & 0 & 1 \\ \midrule
a  &   0.13537  &  0.00364 & n   &  0.00035  &  0.11429\\
b  &  0.00023  &  0.02307 & o  &  0.13081  &  0.00143\\
c  &  0.00039  &  0.05605 & p  &  0.00073  &  0.03637\\
d  &  0.00025  &  0.06873 & q  &  0.00019  &  0.00134\\
e  &  0.21176  &  0.00223 & r  &  0.00041  &  0.10128\\
f   &  0.00018  &  0.03556 & s  &  0.00032  &  0.11069\\
g  &  0.00041  &  0.02751 & t  &  0.00158  &  0.15238\\
h  &  0.00526  &  0.06808 & u  &   0.04352  &  0.00098\\
i  &  0.12193  &  0.00077 & v  &  0.00019  &  0.01608\\
j  &  0.00014  &  0.00326 & w  &  0.00017  &  0.02301\\
k  &   0.00112  &  0.00759 & x  &  0.00030  &  0.00426\\
l  &  0.00143  &  0.07227 & y  &  0.00028  &  0.02542\\
m  &  0.00027  &  0.03897 & z  &  0.00017  &  0.00100\\ \midrule
 space & 0.34226 & 0.00375 & --- & --- & --- \\ \midrule\midrule 
  \end{tabular}
  }
\end{table} 

Suppose that for a given letter~$\ell$, 
we define its HMM2Vec representation~$V(\ell)$ to be
the corresponding row of the matrix~$B^{\transpose}$ in Table~\ref{tab:initFinal_B}.
Then, for example,
\begin{align}\label{eq:V_aest}
\begin{split}
  V(\ma) &= \rowvecc{0.13537}{0.00364}\ \ \ \ 
  V(\me) = \rowvecc{0.21176}{0.00223}\\
  V(\ms) &= \rowvecc{0.00032}{0.11069}\ \ \ \ 
  V(\mt) = \rowvecc{0.00158}{0.15238}
\end{split}
\end{align}
Next, we consider the distance between these HMM2Vec
representations. Instead of using Euclidean distance, we
measure the cosine similarity.\footnote{Cosine similarity
is not a true metric, since it does not, in general, satisfy the triangle inequality.}

The cosine similarity of vectors~$X$ and~$Y$ is the cosine of the angle between
the two vectors. 
Let~$S(X,Y)$ denote the cosine similarity between vectors~$X$ and~$Y$.
Then for~$X=(X_0,X_1,\ldots,X_{n-1})$ and~$Y=(Y_0,Y_1,\ldots,Y_{n-1})$,
$$
  S(X,Y) = \frac{\displaystyle\sum_{i=0}^{n-1} X_i Y_i}{
  	\sqrt{\displaystyle\sum_{i=0}^{n-1} X_i^2}\sqrt{\displaystyle\sum_{i=0}^{n-1} Y_i^2}}
$$
In general, we have~$-1\leq S(X,Y)\leq 1$, but since our HMM2Vec 
encoding vectors consist of probabilities---and hence are non-negative 
values---in this case, we always have~$0\leq S(X,Y)\leq 1$.

When considering cosine similarity, the length of the vectors is irrelevant,
as we are only considering the angle between vectors. Consequently, we
might want to normalize all vectors to be of length one, say, $\widetilde{X}=X/\|X\|$ 
and~$\widetilde{Y}=Y/\|Y\|$, in which case the cosine similarity simplifies
to the dot product
$$
  S(X,Y) = S(\widetilde{X},\widetilde{Y}) = \displaystyle\sum_{i=0}^{n-1} \widetilde{X}_i \widetilde{Y}_i
$$
Henceforth, we use the notation~$\widetilde{X}$ to indicate a vector~$X$ that 
has been normalized to be of length one.

For the vector encodings in~\eref{eq:V_aest},
we find that for the vowels ``a'' and ``e'', the cosine similarity
is~$S(V(\ma),V(\me))=0.9999$. In contrast, 
the cosine similarity of the vowel ``a'' 
and the consonant ``t'' is~$S(V(\ma),V(\mt))=0.0372$. 
The normalized vectors~$V(\ma)$ and~$V(\mt)$ are illustrated in
Figure~\ref{fig:cSim}. Using the notation in this figure, 
cosine similarity is~$S(V(\ma),V(\mt))=\cos(\theta)$

\begin{figure}[!htb]
  \centering
    \input figures/cosineSim.tex
  \caption{
  Normalized vectors~$\widetilde{V}(\ma)$ and~$\widetilde{V}(\mt)$}\label{fig:cSim}
\end{figure}

These results indicate that our HMM2Vec encodings---which are derived from a trained 
HMM---provide useful information on the similarity (or not) of pairs of letters.
Note that we could obtain a vector encoding of any dimension
by simply training an HMM with the number of hidden states~$N$
equal to the desired vector length.

In our experiments below, we consider HMM2Vec embeddings. However, in this
research, models are trained on opcodes instead of letters, and hence
the embeddings are relative to individual opcodes.

\subsubsection{Logistic Regression}

Logistic regression is used widely for classification problems. 
This relatively simple technique relies on the sigmoid function, which 
is also knows as the logistic function, and hence the name. The
sigmoid function is defined as
$$ 
S(x)=\frac{1}{1+e^{-x}} .
$$ 

\begin{figure}[!htb]
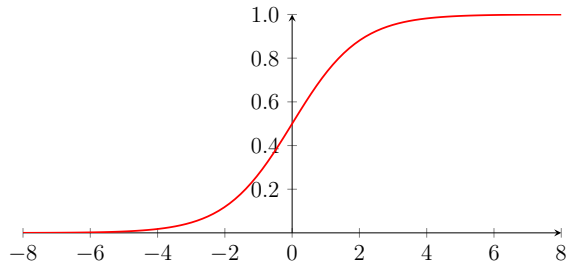

\centering
\input figures/sigmoid.tex
\caption{Graph of sigmoid function}\label{fig:5}
\end{figure}

Logistic regression can be viewed as a modification of linear regression.
As with linear regression, logistic regression 
models the probability that observations take one of two (binary) values. 
Linear regression makes unbounded predictions whereas logistic regression 
converts the probability into the range~0 to~1 due to the use
of the sigmoid function. The graph of the sigmoid function is given in Figure~\ref{fig:5},
from which we can see that the output must be between~0 and~1. 

\section{Experiments and Results}\label{chap:results}

In this section, we discuss our evolution detection experiments and results. 
We divide this section into four subsections, one for each technique considered,
namely, logistic regression, HMM, HMM2Vec and Word2Vec.

\subsection{Logistic Regression Experiments}

\begin{figure}[!htb]
   \centering
   \begin{tabular}{cc}
    \includegraphics[width=0.4\textwidth]{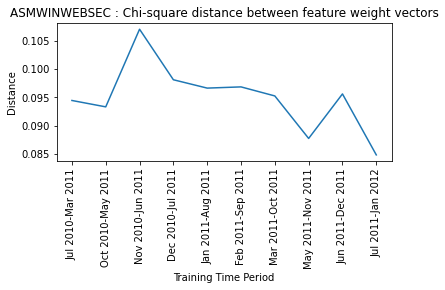}
    & 
    \includegraphics[width=0.4\textwidth]{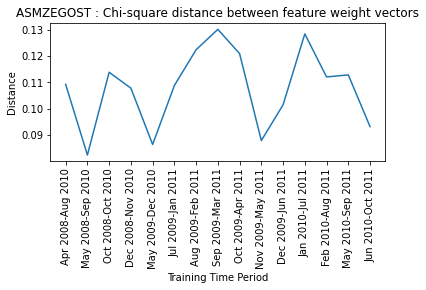}
    \\
    (a) Winwebsec
    &
    (b) Zegost
    \\
    \end{tabular}
    \caption{Logistic regression results}\label{fig:6}
\end{figure}

As mentioned above, in~\cite{paul2021word} the authors use linear SVMs to 
detect potential malware evolution. Logistic regression is a simpler technique that, like SVM,
is widely used for classification. Hence, we train logistic regression models over time-windows,
analogous to the SVM approach in~\cite{paul2021word}. Specifically,
we divide our data into overlapping time windows of one year,
with a slide length of one month. All of the samples from the most recent one year 
time window are taken as the~$+1$ class, while samples from the current month are 
considered as the~$-1$ class, and we train our logistic regression models on the resulting data. 
Each such model is represented by its weights, and we calculate the Euclidian distances 
between these weight vectors to measure the similarity of the models. 
We then plot these distances on a timeline---spikes in the graph indicate that
the model has changed and hence evolution may have occurred. 
Figure~\ref{fig:6} shows the results of our logistic regression experiments for 
Winwebsec and Zegost.

The results in Figure~\ref{fig:6} indicate that we have random fluctuations in the graphs,
rather than significant spikes that would indicate evolution. Although our logistic regression 
model achieves high accuracy in classifying samples, the weights of the hidden layer do 
not appear to provide useful information regarding changes in the malware samples. 
Apparently, the noise inherent in these weights overwhelms the relevant information. 

\subsection{Hidden Markov Model Experiments}

All experiments in this section are based on the top thirty most frequent opcodes
per family, with all other opcodes grouped into a single ``other'' category.
Thus, our HMMs are all based on~$M=31$ distinct symbols. 
We use~$N=2$ hidden states in all experiments.
We conduct two sets of experiments based on hidden Markov models (HMM). 
In both of these approaches, we train models, 
and we then score samples with the resulting models.

For our first set of experiments, we reserve the data from the first one-month
time period to test our models, and hence we do not train a model on this data.
For each subsequent one-month time window, we train a model, and then
score the samples from the first one-month time period versus each of these models.
We refer to this as HMM approach~1.

\begin{figure}[!htb]
\centering
 \begin{tabular}{cc}
    \includegraphics[width=0.42\textwidth]{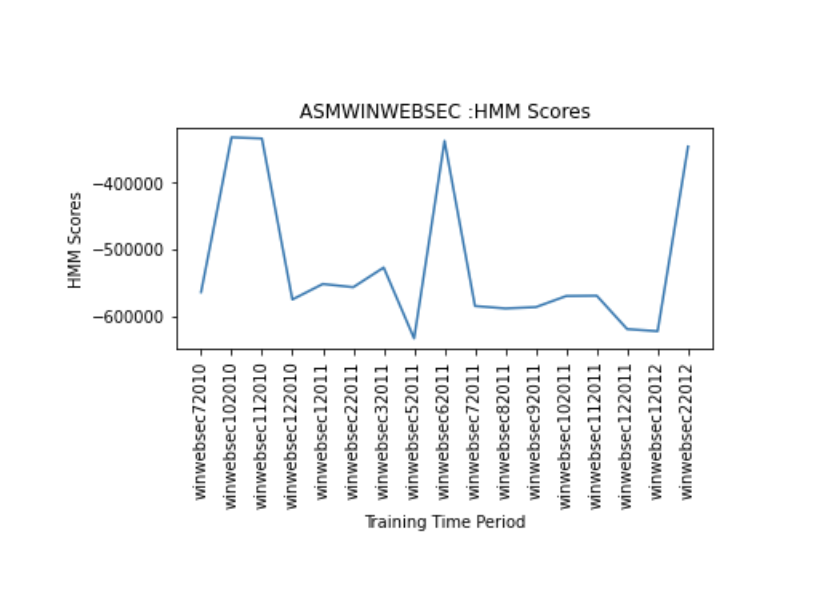}
    & 
    \includegraphics[width=0.4\textwidth]{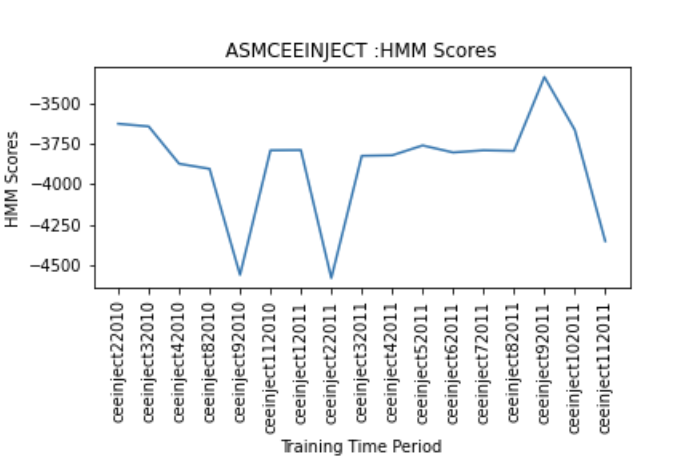}
    \\[-1ex]
    (a) Winwebsec
    &
    (b) CeeInject
    \\
    \multicolumn{2}{c}{\includegraphics[width=0.44\textwidth]{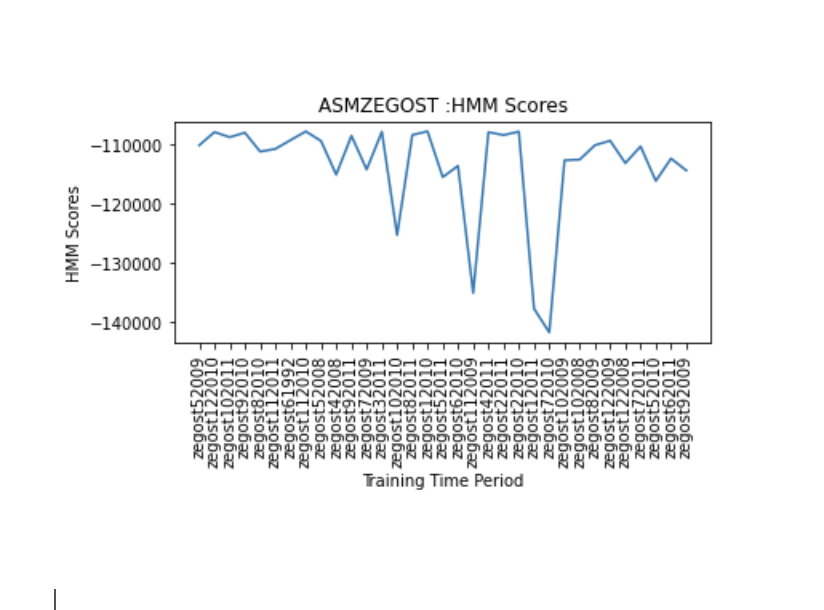}}
    \\[-4ex]
    \multicolumn{2}{c}{(c) Zegost}
    \\
     \end{tabular}
      \caption{HMM approach~1 results for three families}
      \label{fig:7}
\end{figure}

Consider two distinct one-month time periods, say time period~$X$ and~$Y$.
Suppose that we train an HMM on the data from time period~$X$ and another on
the data from time period~$Y$, which we denote as~$\lambda_X$ and~$\lambda_Y$,
respectively. If the samples from~$X$ and~$Y$ are similar, then we expect the 
HMMs~$\lambda_X$ and~$\lambda_Y$ to be similar, and hence they should produce similar
scores on the reserved (first month) data. On the other hand, if the the samples
from time periods~$X$ and~$Y$ differ significantly, then we expect the 
models~$\lambda_X$ and~$\lambda_Y$ to differ, and hence the scores
on the reserved first-month test set should differ significantly.
Figure~\ref{fig:7} shows results for three families based on this 
HMM approach~1.

In Figure~\ref{fig:7}, we observe spikes in the graphs at various points in time, 
with relative stability over extended periods of time. Thus, this approach seems 
to have the potential to detect malware evolution.


Next, we consider another application of HMMs to our data. 
In this case, for each one-month time window,
we use~75\%\ of the available samples for training and reserve~25\%\ for testing. 
Next, we train an HMM for each month---as above, we use~$N=2$, and we have~$M=31$
in each case. 

Suppose we have data from consecutive months that we label as~$X$ and~$Y$.
We train model~$\lambda_{X}$ on the training data from time period~$X$ 
and we train a model~$\lambda_{Y}$ on the training data from time period~$Y$.
We then score each test sample from~$X$ with both~$\lambda_{X}$ 
and model~$\lambda_{Y}$, giving us two score vectors. 
Since an HMM score depends on the length of the observation sequence, and 
since the observation sequence lengths vary between malware samples, 
each scores is normalized by dividing by the length of the observation sequence.
As a result, each score is in the form of a log likelihood per opcode (LLPO).
Note that If we have, say, $m$ test samples in~$X$, 
the score vector obtained from~$\lambda_{X}$ and the score
vector obtained from~$\lambda_{Y}$ will both be of length~$m$.

\begin{figure}[!htb]
   \centering
   \begin{tabular}{cc}
    \includegraphics[width=0.4\textwidth]{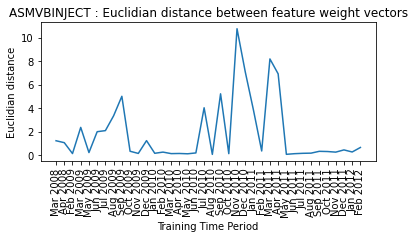}
    & 
    \includegraphics[width=0.4\textwidth]{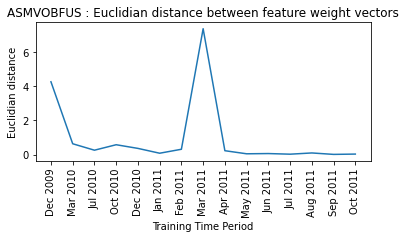}
    \\
    (a) VbInject 
    &
    (b) Vobfus
    \\ \\
    \includegraphics[width=0.4\textwidth]{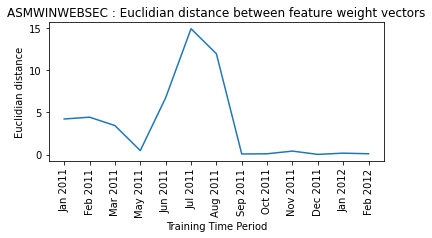}
    &
    \includegraphics[width=0.4\textwidth]{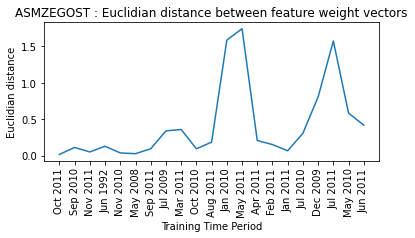}
    \\
    (c) Winwebsec
    &
    (d) Zegost
     \end{tabular}
    \caption{HMM approach~2 results for four families}\label{fig:8}
\end{figure}

Once we generate these two vectors, we compute the Euclidean distance between the vectors,
which we denote as~$d_{X}$. We repeat this scoring process using the test samples
from~$Y$ to obtain a distance~$d_{Y}$, and we define the distance between time
windows~$X$ and~$Y$ to be the average, that is,
$$
  d=\frac{d_{X} + d_{Y}}{2} .
$$
We plot the graph of these distances---small changes 
in the distance from one month to the next suggests minimal change, 
whereas larger distances indicate potential evolution points. 
Figure~\ref{fig:8} gives results for four malware families 
using this HMM-based technique, which we refer to as HMM approach~2.

The results in Figure~\ref{fig:8} indicate that we see significant evolutionary
change points when considering this second HMM technique. Together with the results
for HMM approach~1 in Figure~\ref{fig:7}, these results provide strong evidence 
that HMM-based techniques are a powerful tool for malware evolution detection.

\subsection{HMM2Vec Experiments}

In this section we present our experimental results using HMM2Vec, which
is discussed in Section~\ref{sect:HMM2Vec}. 
In these experiments, we select~$N=2$
and we have~$M=31$. Recall that the HMM2Vec embeddings are determined
by the columns of the~$B$ matrix from our trained HMM,
and that each embedding vector is of length~$N$.

A technical difficulty arises when considering HMM2Vec embeddings. 
That is, the order of the hidden states can vary between models---even 
when training on the same data, different random initializations
can cause the hidden states to differ in the resulting trained models. Since we only
consider models with~$N=2$ hidden states, we account for this possibility in our HMM2Vec 
experiments by computing the distance between~$B$ matrices twice, once with the order of
the rows flipped in one of the models. More precisely, suppose that we 
want to compare the two HMMs~$\lambda=(A,B,\pi)$ 
and~$\widetildeto{N}{\lambda}=(\widetildeto{N}{A},\widetildeto{N}{B},\widetildeto{\pi}{\pi})$,
where~$N=\widetildeto{N}{N}=2$ and~$M=\widetildeto{N}{M}$.
We first compute the distance based on the HMM2Vec embeddings
determined by the matrices~$B$ and~$\widetildeto{N}{B}$
(we ignore~$A$ and~$\widetildeto{N}{A}$,
as well as~$\pi$ and~$\widetildeto{\pi}{\pi}$). Denote the rows of~$B$ 
as~$B_1$ and~$B_2$ and, similarly, 
let~$\widetildeto{N}{B}_{\!1}$ and~$\widetildeto{N}{B}_{\!2}$
be the rows of~$\widetildeto{N}{B}$. Compute
$$
  d_1=d(B_1\| B_2, \widetildeto{N}{B}_{\!1} \|\, \widetildeto{N}{B}_{\!2})
  \mbox{\ and\ }
  d_2=d(B_1\| B_2, \widetildeto{N}{B}_{\!2} \|\, \widetildeto{N}{B}_{\!1})
$$
where~``$\|$'' is the concatenation operator, and~$d(x,y)$ is 
the Euclidean distance between vectors~$x$ and~$y$. 
We define the HMM2Vec
distance between~$\lambda$ and~$\widetildeto{N}{\lambda}$ as
$$
  d(\lambda,\widetildeto{N}{\lambda}) = \min \{d_1,d_2\} .
$$

\begin{figure}[!htb]
   \centering
   \begin{tabular}{cc}
    \includegraphics[width=0.4\textwidth]{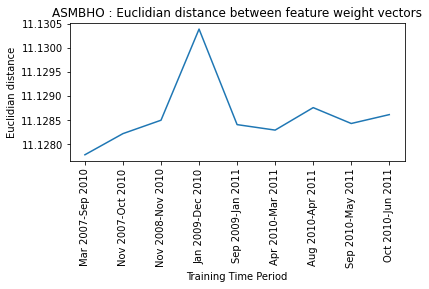}
    & 
    \includegraphics[width=0.4\textwidth]{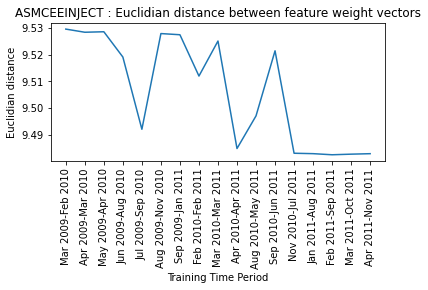}
    \\
    (a) Bho 
    &
    (b) CeeInject
    \\
    \\
    \multicolumn{2}{c}{\includegraphics[width=0.4\textwidth]{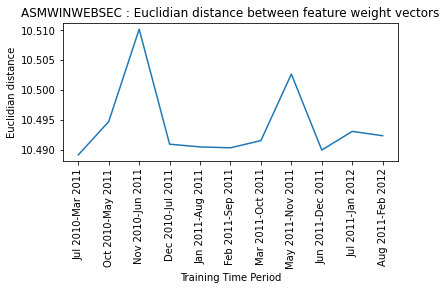}}
    \\
    \multicolumn{2}{c}{(c) Winwebsec}
    \end{tabular}
    \caption{HMM2Vec results for three malware families}\label{fig:9}
\end{figure}

We divide the dataset into overlapping windows of one year, with a slide length of one month
and we train an HMM (with~$N=2$ and~$M=31$)
on each window. We compute the distance between adjacent windows using the 
method described in the previous paragraph, and we graph the resulting
distances. The graphs obtained for three families are given in Figure~\ref{fig:9}.

The results in Figure~\ref{fig:9} indicate that HMM2Vec is successful 
in identifying potential evolution in these particular families. 
We observe significant spikes (i.e., evolutionary points) in most 
families using this technique.

\subsection{Word2Vec Experiments}

In this set of experiments, we use Word2Vec to generate vector embeddings of opcodes.
We compare the resulting models by concatenating the embedding vectors,
and computing the distance between the resulting vectors.
As above, we divide the dataset into overlapping time windows of one year, with a slide length of 
one month. The Word2Vec models are trained as outlined in Section~\ref{sect:w2v}.

\begin{figure}[!htb]
\centering
 \begin{tabular}{cc}
    \includegraphics[width=0.4\textwidth]{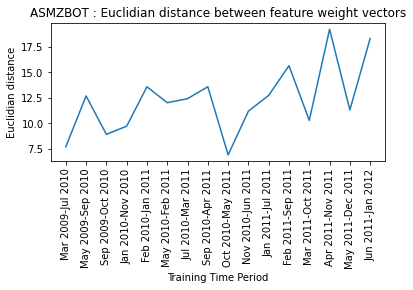}
    & 
    \includegraphics[width=0.4\textwidth]{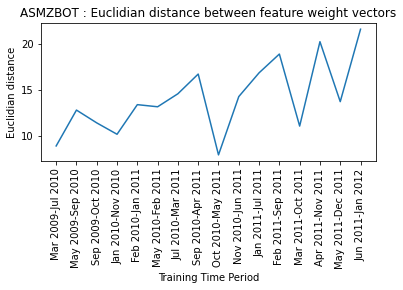}
    \\
    (a) Zbot with~$V=2$
    &
    (b) Zbot with~$V=3$
    \\
    \\
    \multicolumn{2}{c}{\includegraphics[width=0.4\textwidth]{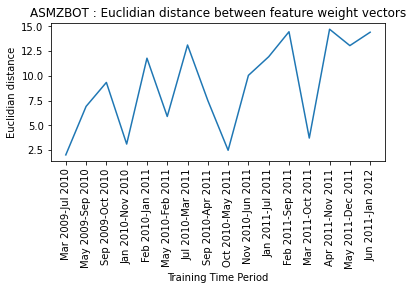}}
    \\
    \multicolumn{2}{c}{(c) Zbot with~$V=5$}
    \end{tabular}
    \caption{Word2Vec with different vector sizes (Zbot family)}\label{fig:10}
\end{figure}

When training Word2Vec, the window size~$W$ refers to the length of the window used
to determine training pairs, while the vector length~$V$ is the number of components
in each embedding vector. We experimented with different window sizes and
found that~$W=5$ works best. We also experimented with different vector 
sizes---in Figure~\ref{fig:10}, we give results for the Zbot family
for~$V=2$, $V=3$, and~$V=5$. In general, 
we do not find any improvement for larger values of~$V$, and
hence we use~$V=2$ 
in all of our subsequent Word2Vec experiments. 

\begin{figure}[!htb]
   \centering
   \begin{tabular}{cc}
    \includegraphics[width=0.4\textwidth]{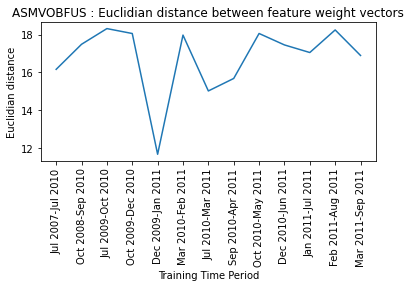}
    &
    \includegraphics[width=0.4\textwidth]{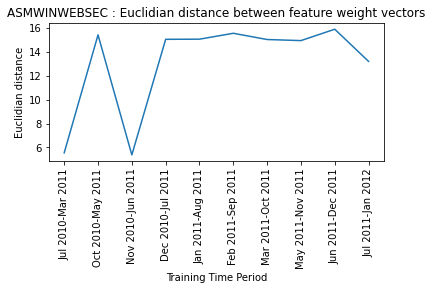}
    \\
    (a) Vobfus
    &  
    (b) Winwebsec
    \\
    \\
    \multicolumn{2}{c}{\includegraphics[width=0.4\textwidth]{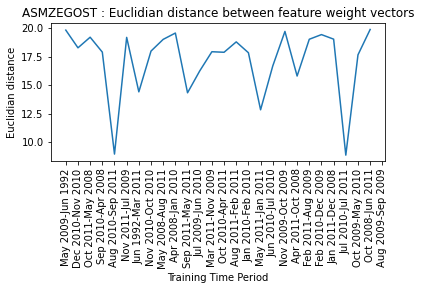}}
    \\
    \multicolumn{2}{c}{(c) Zegost}
    \end{tabular}
    \caption{Word2Vec results for three malware families}\label{fig:11}
\end{figure}

Results from our Word2Vec experiments for three families are given in Figure~\ref{fig:11}.
These results show potential evolutionary points in almost all the malware families
and we conclude that Word2Vec is also a useful technique for detect potential 
malware evolution points.

\subsection{Discussion}

Here, we first discuss the results given by each technique considered in this section.
Then we compare our results to the most closely related previous work.

The two HMM scoring techniques that we first considered provide different models and scores,
yet the results are similar. This provides evidence of 
correctness and consistency, and also some evidence of actual evolution.

Our HMM2Vec and Word2Vec experiments were somewhat different, since they focus on 
longer time windows of one year, whereas the HMM techniques
both are based on one-month time intervals. 
In any case, both HMM2Vec and Word2Vec
performed well and consistently with
each other. Again, this consistency is evidence of correctness
of the implementations, and of evolution detection.

Combining either HMM2Vec or Word2Vec with either of the HMM scoring
techniques provides a two-step strategy for detecting evolution. 
That is, we can use a year-based technique
(either HMM2Vec or Word2Vec) to see if there is any indication of evolution over 
such a time window. If so, we can then use one of the HMM scoring techniques to 
determine where within that one-year window the strongest evolutionary points occur.
In this way, we could rapidly filter out time periods that are unlikely to be of interest,
and then in the secondary phase, detect precise times at which interesting evolutionary
changes have most likely occurred.
For example, both Word2Vec and HMM2Vec indicate that evolution in the
Winwebsec family took place during the time period November~2010--June~2011. 
Then experimenting with the first HMM scoring approach on the Winwebsec family 
during the November~2010--June~2011 time period indicates that 
the precise point of evolution was June~2011.

Next, we compare our work to that in~\cite{paul2021word}, which considered
the same malware evolution problem and used the same dataset as in our research. 
In~\cite{paul2021word}, linear SVM models are trained over on year time windows, with
a slide of one month. The resulting linear SVM model weights are compared using a~$\chi^{2}$
distance computation. Furthermore, Word2Vec feature vectors (derived from opcode
sequences) were used as input features to their SVM models. 
Once a~$\chi^{2}$ similarity graph has been generated, 
an HMM-based approach is used on either sides of a spike
to confirm that evolution has occurred.

Comparing our results with those given in~\cite{paul2021word}, our techniques are
more efficient, as we omit the SVM training and our work factor is less than
their secondary test. In spite of these
simplifications, we find that our detect strategy is at least as sensitive
as that in~\cite{paul2021word}. For example, we see clear 
spikes in some families (e.g., DelfInject, Dorkbot, and Zbot) for which
previous work found, at best, ambiguous results. 

Next we briefly summarize our results per family. We refer to previous 
work in~\cite{paul2021word} in some of these cases.

\begin{description}

\item\textbf{Adload}: For this family, both HMM2Vec and Word2Vec did not 
result in any significant spikes in the graphs, and hence we do not see indications of
evolutionary change. On the other hand, the results given by our HMM techniques 
show significant spikes for this family.

\item\textbf{Bho}: The results generated by Word2Vec for this family 
indicate that malware evolution occurred during 
the September~2009--December~2010 timeframe. 
Using our HMM approach, we are able to see that malware evolution happened 
during October~2010, which is consistent with the Word2Vec results. 
This is also consistent with results given in~\cite{paul2021word}, based on SVM analysis. 


\item\textbf{Bifrose}: The results generated by HMM2Vec did not
provide any major spikes in the graph,
but we can see indications of slower change over time. 
The graph generated by Word2Vec gives us a better understanding of 
changes in this malware family, since we could see that significant evolution 
occurred during the November~2009--March~2011 time period. 
Again, the results given by the HMM approaches narrow down the 
evolution point---in this case, to March 2011. A similar graph is given
in~\cite{paul2021word}, indicating evolution during November~2010--May~2011,
which is consistent with our results. 

\item\textbf{CeeInject}: For CeeInject, we obtain clear results from all experiments we performed. 
The results given by HMM2Vec and Word2Vec shows significant evolution during 
the August~2010--July~2011 time window, and we identify the month of clearest
change as November~2010 based on our HMM approaches. 
The results for this family given in~\cite{paul2021word} show 
similar evolution during September~2010--May~2011.

\item\textbf{DelfInject}: We obtained significant results for this family using 
our HMM-based approaches. This family shows evolution occurring during 
January~2011. In this case, we do not observe significant spikes
for Word2Vec or HMM2vec with their longer time windows. In~\cite{paul2021word},
no evolutionary points are detected for this family.

\item\textbf{Dorkbot}: Similar to CeeInject, we obtain strong results 
on Dorkbot from all of our experiments. Specifically, the evidence strongly points to 
malware evolution during~2011.

\item\textbf{Hupigon}: The results received from Word2Vec technique show 
that significant malware evolution in this family happened during 
the July~2010--April~2011 period. Results from the HMM approaches narrow the time period
to February~2011. Results given by the SVM approach in~\cite{paul2021word} are
consistent with these results.

\item\textbf{Ircbot}: The results generated by Word2Vec indicates that 
malware evolution occurred in this family slowly throughout~2011. That is,
there is no major spikes observed, but the graph shows a slow changing trend.

\item\textbf{Obfuscator}: We could not derive significant information from this family. 
Graphs plotted on this family had many spikes which we could not interpret 
regarding malware evolution.

\item\textbf{Rbot}: Graphs generated based on Word2Vec show significant evolution 
in this malware family. Significant results were not observed for this family 
in any previous research.

\item\textbf{VbInject}: We could not observe a significant spike in this malware family 
in any of our experiments. .

\item\textbf{Vobfus}: The results generated by our experiments shows that 
evolution in this family occurred during the December~2009--January~2011 timeframe. 
The results given in~\cite{paul2021word} indicate evolution during November~2010--May~2011. 

\item\textbf{Winwebsec}: We observe evolution in this malware family using 
Word2Vec, where a spike appears in December~2010--July~2011. The
previous research in~\cite{paul2021word} did not indicate evolution for this family.

\item\textbf{Zbot}: Experiments conducted on this family inidcate significant changes. Specifically,
we observe a spike between April~2011--November~2011. 

\item\textbf{Zegost}: From our Word2Vec experiments, we see significant spikes 
in the August~2010--September~2011 and July~2010--July~2011 timeframes. 

\end{description}

Our experiments indicate significant evolution in almost all the malware families
considered. By comparing the results given by our two HMM techniques, 
HMM2Vec, and Word2Vec, we can see that there are clear similarities in 
the results for most families. When we observe such similar evolution points
across different experiments, it increases our confidence in the results.
As further evidence, we found that the evolutionary points generated 
in previous research in~\cite{paul2021word} matches with our experiments,
and we detect additional points of interested, as compared to previous research,
indicating that our techniques may be somewhat more sensitive.

In some cases, we found potential evolutionary points with the HMM techniques, but not with
HMM2Vec or Word2Vec. We conjecture that this is a result of the longer time windows (one year)
used in the latter two approaches, while the HMM techniques are based on monthly time windows.
These longer time windows may not be as sensitive in cases where a changes are 
less pronounced or transient.

\section{Conclusion And Future Work}
\label{chap:conclusion}

In previous research---first in~\cite{wadkar2020detecting} and subsequently 
in~\cite{paul2021word}---it 
has been shown that malware evolution can be detected using
machine learning techniques. In this paper, we extend this previous work
by exploring additional learning techniques. We find that various HMM-based
techniques and Word2Vec provide powerful tools for automatically
detecting malware evolution.

Here, we conducted all of our experiments based on mnemonic opcodes derived 
from the malware samples. For future work, it would be useful to consider experiments 
with other features extracted from the malware samples. While mnemonic opcodes
perform well, extracting such opcodes is relatively expensive. It is possible that 
other, less costly features can be used.
Also, by considering dynamic features, we might gain more information 
about evolution within a malware family. 
Finally, the use of additional neural networking and deep learning techniques 
should be considered. Word2Vec performed well, and it is likely that
more sophisticated techniques would result in more discriminative
ability, which would enable more fine grained analysis of evolutionary trends.

\bibliographystyle{plain}
\bibliography{references.bib}

\end{document}

%% file: preamble.tex
\usepackage{amsmath,amsthm,mathtools,amsfonts,amssymb,amsxtra,amsopn}
\usepackage{pgfplots}
\usepgfplotslibrary{colormaps}
\pgfplotsset{compat=1.15}
\usepackage{pgfplotstable}
\usetikzlibrary{pgfplots.statistics} 
\usepackage{filecontents}
\usepackage{graphicx}
\usepackage{multirow}
\usepackage{tabu}
\usepackage{algorithmic}
\usepackage{longtable}
\usepackage{booktabs}
\usepackage{listings}
\usepackage{cmap}
\usepackage{colortbl}
\usepackage{adjustbox}
\usepackage{epsfig}
\usepackage{xstring}

\usepackage{subfigure}

\PassOptionsToPackage{hyphens}{url}

\usepackage{hyperref}
\hypersetup{colorlinks=true,linkcolor=black,citecolor=black,urlcolor=blue,filecolor=black}
\hypersetup{pdfpagemode=UseNone,pdfstartview=}

\usepackage[tableposition=top]{caption}
\usepackage{enumitem}
\setlist[itemize]{noitemsep, topsep=0pt}

\advance\oddsidemargin by -0.45in
\advance\textwidth by 0.9in

\advance\topmargin by -0.3in
\advance\textheight by 0.6in

\def\eref#1{(\ref{#1})}

\long\def\symbolfootnotetext[#1]#2{\begingroup%
\def\thefootnote{\fnsymbol{footnote}}\footnotetext[#1]{#2}\endgroup}

\DeclareMathOperator{\ma}{a}
\DeclareMathOperator{\me}{e}
\DeclareMathOperator{\ms}{s}
\DeclareMathOperator{\mt}{t}

\def\z{\phantom{0}}

\def\transpose{\raisebox{2pt}{$\scriptstyle\intercal$}}

\pgfkeys{
    /pgf/number format/fixed zerofill=true }
    
\pgfplotstableset{
    color cells/.code={%
        \pgfqkeys{/color cells}{#1}%
        \pgfkeysalso{%
            postproc cell content/.code={%
                \begingroup
                \pgfkeysgetvalue{/pgfplots/table/@preprocessed cell content}\value
\ifx\value\empty
\endgroup
\else
                \pgfmathfloatparsenumber{\value}%
                \pgfmathfloattofixed{\pgfmathresult}%
                \let\value=\pgfmathresult
                \pgfplotscolormapaccess
                    [\pgfkeysvalueof{/color cells/min}:\pgfkeysvalueof{/color cells/max}]%
                    {\value}%
                    {\pgfkeysvalueof{/pgfplots/colormap name}}%
                \pgfkeysgetvalue{/pgfplots/table/@cell content}\typesetvalue
                \pgfkeysgetvalue{/color cells/textcolor}\textcolorvalue
                \toks0=\expandafter{\typesetvalue}%
                \xdef\temp{%
                    \noexpand\pgfkeysalso{%
                        @cell content={%
                            \noexpand\cellcolor[rgb]{\pgfmathresult}%
                            \noexpand\definecolor{mapped color}{rgb}{\pgfmathresult}%
                            \ifx\textcolorvalue\empty
                            \else
                                \noexpand\color{\textcolorvalue}%
                            \fi
                            \the\toks0 %
                        }%
                    }%
                }%
                \endgroup
                \temp
\fi
            }%
        }%
    }
}

\def\OO{{\cal O}}

\makeatletter
\def\mathcenterto#1#2{\mathclap{\phantom{#1}\mathclap{#2}}\phantom{#1}}
\let\old@widetilde\widetilde
\def\widetildeto#1#2{\mathcenterto{#2}{\old@widetilde{\mathcenterto{#1}{#2\,}}}}
\let\old@widehat\widehat
\def\widehatto#1#2{\mathcenterto{#2}{\old@widehat{\mathcenterto{#1}{#2\,}}}}
\makeatother

\def\rowvecc#1#2{\left(\!\begin{array}{cc} #1 & #2\end{array}\!\right)}
\def\srowvecc#1#2{(\!\begin{array}{cc} 
      \noexpandarg\IfBeginWith{#1}{-}{\! #1}{#1}
    & #2\kern-0.5pt\end{array}\!)}
\def\rowvecc#1#2{\left(\!\begin{array}{cc} 
      \noexpandarg\IfBeginWith{#1}{-}{\! #1}{#1}
    & #2\kern-0.5pt\end{array}\!\right)}
\def\rowveccc#1#2#3{\left(\!\begin{array}{ccc} 
      \noexpandarg\IfBeginWith{#1}{-}{\! #1}{#1}
    & #2 
    & #3\kern-0.5pt\end{array}\!\right)}
\def\rowvecccc#1#2#3#4{\left(\!\begin{array}{cccc}
      \noexpandarg\IfBeginWith{#1}{-}{\! #1}{#1}
    & #2 
    & #3 
    & #4\kern-0.5pt\end{array}\!\right)}
\def\srowvecccc#1#2#3#4{\bigl(\!\begin{array}{cccc}
      \noexpandarg\IfBeginWith{#1}{-}{\! #1}{#1}
    & #2 
    & #3 
    & #4\kern-0.5pt\end{array}\!\bigr)}
\def\rowveccccc#1#2#3#4#5{\left(\!\begin{array}{ccccc} 
      \noexpandarg\IfBeginWith{#1}{-}{\! #1}{#1}
    & #2
    & #3
    & #4
    & #5\kern-0.5pt\end{array}\!\right)}
\def\srowvecccccc#1#2#3#4#5#6{(\!\begin{array}{cccccc} 
      \noexpandarg\IfBeginWith{#1}{-}{\! #1}{#1}
    & #2
    & #3
    & #4
    & #5
    & #6\kern-0.5pt\end{array}\!)}
\def\rowvecccccc#1#2#3#4#5#6{\left(\!\begin{array}{cccccc} 
      \noexpandarg\IfBeginWith{#1}{-}{\! #1}{#1}
    & #2
    & #3
    & #4
    & #5
    & #6\kern-0.5pt\end{array}\!\right)}

%% file: figures/hmm.tex
\begin{tikzpicture}[scale=0.8,every node/.style={scale=0.8}]
    
    \draw[thick,color=blue] (0,0) rectangle (1,1);
    \draw[thick,color=blue] (2.5,0) rectangle (3.5,1);
    \draw[thick,color=blue] (5,0) rectangle (6,1);
    \draw[thick,color=blue] (10,0) rectangle (11,1);

    \draw[thick,color=green] (0.5,4.5) circle (0.575);
    \draw[thick,color=green] (3,4.5) circle (0.575);
    \draw[thick,color=green] (5.5,4.5) circle (0.575);
    \draw[thick,color=green] (10.5,4.5) circle (0.575);
    
    \node at (0.5,0.5){$\OO_0$};
    \node at (3,0.5){$\OO_1$};
    \node at (5.5,0.5){$\OO_2$};
    \node at (8,0.5){$\cdots$};
    \node at (10.5,0.5){$\OO_{T-1}$};

    \node at (0.5,4.5){$X_0$};
    \node at (3,4.5){$X_1$};
    \node at (5.5,4.5){$X_2$};
    \node at (8,4.5){$\cdots$};
    \node at (10.5,4.5){$X_{T-1}$};
       
    \node at (1.7,4.8){$A$};
    \node at (4.2,4.8){$A$};
    \node at (6.7,4.8){$A$};
    \node at (9.2,4.8){$A$};
    
    \node at (0.2,2.1){$B$};
    \node at (2.7,2.1){$B$};
    \node at (5.2,2.1){$B$};
    \node at (10.2,2.1){$B$};
    
     \draw[thick,color=black,->] (1.075,4.5) -- (2.425,4.5);
     \draw[thick,color=black,->] (3.575,4.5) -- (4.925,4.5);
     \draw[thick,color=black,->] (6.075,4.5) -- (7.425,4.5);
     \draw[thick,color=black,->] (8.575,4.5) -- (9.925,4.5);

     \draw[thick,color=black,->] (0.5,3.925) -- (0.5,1);
     \draw[thick,color=black,->] (3.0,3.925) -- (3.0,1);
     \draw[thick,color=black,->] (5.5,3.925) -- (5.5,1);
     \draw[thick,color=black,->] (10.5,3.925) -- (10.5,1);

    \draw[thick,dashed,color=red] (-0.3,3) -- (11.2,3);
   
\end{tikzpicture}

%% file: figures/w2v.tex
\begin{tikzpicture}[scale=0.9]
    
    \draw[thick,color=green] (4.25,8.5) circle (0.575);

    \draw[thick,color=blue] (2.5,5.5) rectangle (3.5,6.5);
    \node at (4.25,6.0) {$\cdots$};
    \draw[thick,color=blue] (5.0,5.5) rectangle (6.0,6.5);
        
    \draw[thick,color=red,rotate around={45:(1.75,3.0)}] (1.25,2.5) rectangle (2.25,3.5);
    \draw[thick,color=red,rotate around={45:(6.75,3.0)}] (6.25,2.5) rectangle (7.25,3.5);

    \draw[thick,color=green,->] (4.0,7.97) -- (3.0,6.52);
    \draw[thick,color=green,->] (4.175,7.92) -- (3.75,6.52);
    \draw[thick,color=green,->] (4.325,7.92) -- (4.75,6.52);
    \draw[thick,color=green,->] (4.5,7.97) -- (5.5,6.52);
    
    \draw[thick,color=blue,->] (2.7,5.5) -- (2.0,3.52);
    \draw[thick,color=blue,->] (2.9,5.5) -- (3.25,3.5);
    \draw[thick,color=blue,->] (3.1,5.5) -- (4.75,3.51);
    \draw[thick,color=blue,->] (3.3,5.5) -- (6.3,3.3);

    \draw[thick,color=blue,->] (5.2,5.5) -- (2.2,3.3);
    \draw[thick,color=blue,->] (5.4,5.5) -- (3.75,3.51);
    \draw[thick,color=blue,->] (5.6,5.5) -- (5.25,3.5);
    \draw[thick,color=blue,->] (5.8,5.5) -- (6.5,3.52);

    \draw[thick,color=red,->] (1.75,2.3) -- (1.75,1.3);
    \draw[thick,color=red,->] (6.75,2.3) -- (6.75,1.3);

    \node at (4.25,8.5) {$V$};

    \node at (3.0,6.0) {$\ell_0$};
    \node at (5.5,6.0) {$\ell_{\kern-1pt N-1}$};

    \node at (1.75,3.0) {$\omega_0$};
    \node at (4.25,3.0) {$\cdots$};
    \node at (6.75,3.0) {$\omega_{\kern-1pt M-1}$};

    \node at (1.75,0.85) {$p_0$};
    \node at (4.25,0.85) {$\cdots$};
    \node at (7.1,0.85) {$p_{M-1}$};
    
    \node at (9.5,8.5) {Input vector};
    \node at (9.5,6.0) {Hidden layer};
    \node at (9.5,3.0) {Output layer};
    \node at (9.5,0.85) {Probability};

\end{tikzpicture}

%% file: figures/cosineSim.tex
\begin{tikzpicture}[scale=1.75,every node/.style={scale=1.0}]

    \draw[thick,color=black] (0.0,0.0) circle(1.0);

    \draw [black,thick,domain=2:90,->] plot ({0.33*cos(\x)}, {0.33*sin(\x)});
   
    \draw[thick,color=red,->] (0.0,0.0) -- (0.9996,0.0269);
    \node at (1.3,0.02) {$\widetilde{V}(\ma)$};

    \draw[thick,color=blue,->] (0.0,0.0) -- (0.010368,0.999946);
    \node at (0.125,1.15) {$\widetilde{V}(\mt)$};

    \node at (0.3,0.33) {$\theta$};



\end{tikzpicture}

%% file: figures/sigmoid.tex
\begin{tikzpicture}[scale=0.85,every node/.style={scale=0.85}]
    \begin{axis}%
    [
        height=5cm,
        width=10cm,
        xmin=-8,
        xmax=8,
        axis x line=bottom,
        ytick={0.0,0.2,0.4,0.6,0.8,1.0},
        ymax=1,
        axis y line=middle,
        x tick label style={
    	    /pgf/number format/.cd,
	    1000 sep={},
    	    fixed,
    	    fixed zerofill,
    	    precision=0},
        y tick label style={
    	    /pgf/number format/.cd,
	    1000 sep={},
    	    fixed,
    	    fixed zerofill,
    	    precision=1}
   ]
        \addplot%
        [
            red,thick,
            mark=none,
            samples=100,
            domain=-8:8,
        ]
        (x,{1/(1+exp(-x))});
    \end{axis}
\end{tikzpicture}

%% file: lolitha.bbl
\begin{thebibliography}{10}

\bibitem{aycock2006computer}
John Aycock.
\newblock {\em Computer Viruses and Malware}.
\newblock Springer, 2006.

\bibitem{BaratMarius2013Asoc}
Marius Barat, Dumitru-Bogdan Prelipcean, and Dragoş Gavriluţ.
\newblock A study on common malware families evolution in 2012.
\newblock {\em Journal of Computer Virology and Hacking Techniques},
  9(4):171--178, 2013.

\bibitem{borello2008code}
Jean-Marie Borello and Ludovic M{\'e}.
\newblock Code obfuscation techniques for metamorphic viruses.
\newblock {\em Journal in Computer Virology}, 4(3):211--220, 2008.

\bibitem{BrownCorpus}
The {B}rown corpus of standard {A}merican {E}nglish.
\newblock \url{http://www.cs.toronto.edu/~gpenn/csc401/a1res.html}.

\bibitem{ChenZhongqiang2012Mcat}
Zhongqiang Chen, Mema Roussopoulos, Zhanyan Liang, Yuan Zhang, Zhongrong Chen,
  and Alex Delis.
\newblock Malware characteristics and threats on the {I}nternet ecosystem.
\newblock {\em The Journal of Systems \&\ Software}, 85(7):1650--1672, 2012.

\bibitem{DamodaranAnusha2017Acos}
Anusha Damodaran, Fabio Troia, Corrado Visaggio, Thomas Austin, and Mark Stamp.
\newblock A comparison of static, dynamic, and hybrid analysis for malware
  detection.
\newblock {\em Journal of Computer Virology and Hacking Techniques},
  13(1):1--12, 2017.

\bibitem{GuptaA2009Aeso}
A~Gupta, P~Kuppili, A~Akella, and P~Barford.
\newblock An empirical study of malware evolution.
\newblock In {\em First International Communication Systems and Networks and
  Workshops}, pages 1--10, 2009.

\bibitem{gupta2009empirical}
Archit Gupta, Pavan Kuppili, Aditya Akella, and Paul Barford.
\newblock An empirical study of malware evolution.
\newblock In {\em First International Communication Systems and Networks and
  Workshops}, pages 1--10, 2009.

\bibitem{KimSamuel2018PHAf}
Samuel Kim.
\newblock {PE} header analysis for malware detection.
\newblock Master's thesis, San Jose State University, Department of Computer
  Science, 2018.

\bibitem{ma2006finding}
Justin Ma, John Dunagan, Helen~J Wang, Stefan Savage, and Geoffrey~M Voelker.
\newblock Finding diversity in remote code injection exploits.
\newblock In {\em Proceedings of the 6th ACM SIGCOMM Conference on Internet
  Measurement}, pages 53--64, 2006.

\bibitem{koob}
Robert McMillan.
\newblock {COMPUTERWORLD}: Researchers take down {K}oobface servers ---
  {C}riminals behind the botnet made more than \$2 million in one year, 2010.
\newblock
  \url{https://www.computerworld.com/article/2750985/researchers-take-down-koobface-servers.html}.

\bibitem{MercaldoFrancesco2018Aeso}
Francesco Mercaldo, Andrea Di~Sorbo, Corrado~Aaron Visaggio, Aniello Cimitile,
  and Fabio Martinelli.
\newblock An exploratory study on the evolution of {A}ndroid malware quality.
\newblock {\em Journal of Software: Evolution and Process}, 30(11), 2018.

\bibitem{Rbot}
Microsoft.
\newblock Win32 {Rbot} detected with {W}indows {D}efender antivirus.
\newblock
  \url{https://www.microsoft.com/en-us/wdsi/threats/malware-encyclopedia
  -description?Name=Win32%2FRbot}, 2005.

\bibitem{Hotbar}
Microsoft.
\newblock Adware: Win32 {Hotbar} detected with {W}indows {D}efender antivirus.
\newblock \url{https://www
  .microsoft.com/en-us/wdsi/threats/malware-encyclopedia
  -description?Name=Adware%3AWin32%2FHotbar}, 2006.

\bibitem{Hupigon}
Microsoft.
\newblock Backdoor: Win32 {Hupigon} detected with {W}indows {D}efender
  antivirus.
\newblock \url{https://www
  .microsoft.com/en-us/wdsi/threats/malware-encyclopedia
  -description?Name=Backdoor%3AWin32%2FHupigon}, 2006.

\bibitem{CeeInject}
Microsoft.
\newblock Virtool: Win32 {CeeInject} detected with {W}indows {D}efender
  antivirus.
\newblock \url{https://www
  .microsoft.com/en-us/wdsi/threats/malware-encyclopedia
  -description?Name=VirTool%3AWin32%2FCeeInject}, 2007.

\bibitem{DelfInject}
Microsoft.
\newblock Virtool: Win32 {DelfInject} detected with {W}indows {D}efender
  antivirus.
\newblock \url{https://www.microsoft.com/en-us/wdsi/threats/
  malware-encyclopedia-description?Name=VirTool:Win32/
  DelfInject&ThreatID=-2147369465}, 2007.

\bibitem{VBinject}
Microsoft.
\newblock Virtool: Win32 {VBInject} detected with {W}indows {D}efender
  antivirus.
\newblock \url{https://www
  .microsoft.com/en-us/wdsi/threats/malware-encyclopedia
  -description?Name=VirTool:Win32/VBInject&ThreatID= -2147367171}, 2010.

\bibitem{Vobfus}
Microsoft.
\newblock Win32 {Vobfus} detected with {W}indows {D}efender antivirus.
\newblock \url{https://www
  .microsoft.com/en-us/wdsi/threats/malware-encyclopedia
  -description?name=win32%2Fvobfus}, 2010.

\bibitem{Winwebsec}
Microsoft.
\newblock Win32 {Winwebsec} detected with {W}indows {D}efender antivirus.
\newblock \url{https:// www.microsoft.com/security/portal/threat/encyclopedia/
  entry.aspx?Name=Win32%2fWinwebsec}, 2010.

\bibitem{Obfuscator}
Microsoft.
\newblock Win32 {Obfuscator} detected with {W}indows {D}efender antivirus.
\newblock \url{https://www
  .microsoft.com/en-us/wdsi/threats/malware-encyclopedia
  -description?Name=Win32%2FObfuscator}, 2011.

\bibitem{Zbot}
Microsoft.
\newblock Win32 {Zbot} detected with {W}indows {D}efender antivirus.
\newblock \url{http://www.symantec.com/ security
  response/writeup.jsp?docid=2010-011016-3514 -99}, 2011.

\bibitem{Zegost}
Microsoft.
\newblock Win32 {Zegost} detected with {W}indows {D}efender antivirus.
\newblock \url{https://www.symantec
  .com/security-center/writeup/2011-060215-2826-99}, 2011.

\bibitem{Dorkbot}
Microsoft.
\newblock Worm: Win32 {Dorkbot} detected with {W}indows {D}efender antivirus.
\newblock \url{https://www
  .microsoft.com/en-us/wdsi/threats/malware-encyclopedia
  -description?Name=Worm%3AWin32/Dorkbot}, 2011.

\bibitem{Bifrose}
Microsoft.
\newblock Win32 {Bifrose} detected with {W}indows {D}efender antivirus.
\newblock \url{https://www.trendmicro.com/vinfo/
  us/threat-encyclopedia/malware/bifrose}, 2012.

\bibitem{w2v}
Tomas Mikolov, Kai Chen, Greg Corrado, and Jeffrey Dean.
\newblock Efficient estimation of word representations in vector space.
\newblock \url{https://arxiv.org/abs/1301.3781}, 2013.

\bibitem{w2v2}
Tomas Mikolov, Ilya Sutskever, Kai Chen, Greg Corrado, and Jeffrey Dean.
\newblock Distributed representations of words and phrases and their
  compositionality.
\newblock
  \url{https://papers.nips.cc/paper/5021-distributed-representations-of-words-and-phrases-and-their-compositionality.pdf},
  2013.

\bibitem{nappa2015malicia}
Antonio Nappa, M~Zubair Rafique, and Juan Caballero.
\newblock The {MALICIA} dataset: Identification and analysis of drive-by
  download operations.
\newblock {\em International Journal of Information Security}, 14(1):15--33,
  2015.

\bibitem{ouellette2013countering}
Jacob Ouellette, Avi Pfeffer, and Arun Lakhotia.
\newblock Countering malware evolution using cloud-based learning.
\newblock In {\em 8th International Conference on Malicious and Unwanted
  Software}, pages 85--94, 2013.

\bibitem{paul2021word}
Sunhera Paul and Mark Stamp.
\newblock Word embedding techniques for malware evolution detection.
\newblock In {\em Malware Analysis Using Artificial Intelligence and Deep
  Learning}, pages 321--343. Springer, 2021.

\bibitem{LRR}
Lawrence~R. Rabiner.
\newblock A tutorial on hidden {M}arkov models and selected applications in
  speech recognition.
\newblock {\em Proceedings of the IEEE}, 77(2):257--286, 1989.
\newblock \url{https://www.cs.sjsu.edu/~stamp/RUA/Rabiner.pdf}.

\bibitem{rezaei2016malware}
Saeid Rezaei, Ali Afraz, Fereidoon Rezaei, and Mohammad~Reza Shamani.
\newblock Malware detection using opcodes statistical features.
\newblock In {\em 8th International Symposium on Telecommunications}, IST,
  pages 151--155, 2016.

\bibitem{stamp2011information}
Mark Stamp.
\newblock {\em Information Security: Principles and Practice}.
\newblock Wiley, 2011.

\bibitem{StampMark2018Itml}
Mark Stamp.
\newblock {\em Introduction to Machine Learning with Applications in
  Information Security}.
\newblock CRC Press, Boca Raton, 2018.

\bibitem{MarkStamp}
Mark Stamp.
\newblock A revealing introduction to hidden {M}arkov models.
\newblock \url{https://www.cs.sjsu.edu/~stamp/RUA/HMM.pdf}, 2018.

\bibitem{wadkar2020detecting}
Mayuri Wadkar, Fabio Di~Troia, and Mark Stamp.
\newblock Detecting malware evolution using support vector machines.
\newblock {\em Expert Systems with Applications}, 143, 2020.

\end{thebibliography}
